\documentclass[letter]{aa} 

\usepackage{graphicx}
\usepackage{txfonts}
\usepackage{amssymb}
\begin{document}
   \title{The H-test probability distribution revisited: Improved sensitivity}

   \subtitle{}

   \author{O.C. de Jager\inst{1,3}
          \and
          I. B\"usching\inst{1,2}
          }

   \offprints{O.C. de Jager}

   \institute{Unit for Space Physics, School of
  Physics, North-West University, 2520 Potchefstroom, South Africa\\
              \email{okkie.dejager@nwu.ac.za}
         \and
             Centre for High Performance Computing (CHPC), CSIR Campus, 15 Lower Hope St. Rosebank, Cape Town, South Africa\\
	\and
	     South African Department of Science \& Technology and National Research Foundation
		Reseach Chair: Astrophysics \& Space Science\\
             }

   \date{Submitted to A\&A Letters}


  \abstract
 {}

   {Aims: To provide a significantly improved probability distribution for the H-test
for periodicity in X-ray and $\gamma$-ray arrival times, which is already extensively used by the $\gamma$-ray pulsar
community. Also, to obtain an analytical 
probability distribution for stacked test statistics in the case of a search for 
pulsed emission from an ensemble of pulsars where the significance per pulsar is relatively
low, making individual detections insignificant on their own. This information is timely given the recent rapid discovery of new pulsars with the Fermi-LAT t $\gamma$-ray telescope. 
}
   {Methods: Approximately $10^{14}$ realisations of the H-statistic ($H$) for random (white) 
noise is calculated from a random number generator for which the repitition cycle is $\gg 10^{14}$.
From these numbers the probability distribution $P(>H)$ is calculated.  
}
   {Results: The distribution of $H$ is is found to be exponential with parameter $\lambda=0.4$
so that the cumulative probability distribution $P(>H)=\exp{(-\lambda H)}$. If we stack independent values for $H$, the sum of $K$ such values would follow the Erlang-K distribution with parameter $\lambda$ for which the cumulative probability distribution is also a simple analytical expression. 
}
   {Conclusion: 

Searches for weak pulsars with unknown pulse profile shapes in the Fermi-LAT, Agile or other X-ray data bases should benefit from the {\it H-test} 
since it is known to be powerful against a broad range of pulse profiles, which introduces only a single statistical trial if only the {\it H-test} is used. The new probability distribution presented here favours the detection of weaker pulsars
in terms of an improved sensitivity relative to the previously known distribution.
}

\titlerunning{Improved distribution of the H-statistic}
   \maketitle
%

\section{Introduction}
When searching for a periodic signal in X-ray or $\gamma$-ray arrival times dominated by noise, we may
either perform a blind search for $\gamma$-ray pulsars as demonstrated by the Fermi-LAT Collaboration (Abdo et al. \cite{fermi1}), or,
search for such a signal where the frequency parameters have been prescribed by contemporary
radio data (Weltevrede \cite{fermi2}). Following the folding of say $N$ arrival times $t_1, ...,t_N$ modulo the pulsar spin parameters, we
arrive at a set of phases $\theta_i$, $i=1, ...N$. 
However, in the case of blind searches, Atwood et al. \cite{atwood} introduced a time differencing
technique in which case the number of trial periods is significantly reduced.

de Jager, Swanepoel \& Raubenheimer (\cite{dsr}, hereafter DSR) reviewed the general class of {\it Beran} statistics (Beran \cite{beran}), from which the most general test statistics such as 
Pearson's $\chi^2$, Rayleigh and $Z^2_m$ statistics are derived, and from within this class they derived the
well known {\it H-test} for X-ray and $\gamma$-ray Astronomy. 

The probability distribution of the {\it H-test} statistic as given by DSR 
was derived from Monte Carlo simulations employing $\sim 10^8$ simulations. The computational power and random number
simulators on typical IBM machines during the 1980's had limited ranges of applicability and the {\it H-test} suffered
accordingly. For values of $H<23$ we found that the probability distribution was exponential with parameter $\lambda=0.398$ (or 0.4),
whereas a hard tail developed for $H>23$, which resulted in a significant compromise in sensitivity.

The old version of the {\it H-test} probability distribution is already extensively used by e.g. the Fermi-LAT Collaboration 
for pulsar searches (e.g. Abdo et al. \cite{fermi1} and Weltevrede et al. \cite{fermi2}), and
from this paper it will become clear that the significances assigned to pulsar detections (or non-detections) may be too
conservative, so that some pulsars may be missed, especially when many trial periods are involved, so that large values
of the $H$-statistic are required for a significant detection
.
In this Letter we notify the community that all previous published significances from the {\it H-test} should be reassessed, based
on the new improved distribution presented below. Before we do so, we first briefly review the origin and properties of the {\it H-test}.

\section{The Beran class of test statistics - towards the H-test}

Let $\theta$ be the pulsar phase measured on the interval $[0,2\pi)$, so that a full rotation corresponds to $2\pi$.
Assuming noise (e.g. from cosmic rays) are also present such that the pulsed fraction is $p\le 1$. Let $f_s(\theta)$ be the observed line-of-sight
pulse profile in the absence of noise. The case $p=0$ then corresponds to no signal (pure noise), whereas $p=1$ corresponds to no noise (pure pulsed signal). The observed light curve $f(\theta)$ can then be represented as a mixing of the noise and signal distributions
(see Eqn (2) of DSR \cite{dsr})
\begin{equation}
 f(\theta)=pf_s(\theta)+\frac{(1-p)}{2\pi}
\end{equation}
The general form of the Beran statistic is given by Beran (\cite{beran}) in the form (see also DSR \cite{dsr})
\begin{equation}
\label{beran}
 \psi(f)=\int_0^{2\pi}(f(\theta)-\frac{1}{2\pi})^2 d\theta
=p^2\int_0^{2\pi}(f_s(\theta)-\frac{1}{2\pi})^2 d\theta ,
\end{equation}
It is thus clear that the {\it Beran} statistic measures the integrated squared distance between the
pulse profile and uniformity, so that if $p \rightarrow 0$, then $\psi \rightarrow 0$ as well, or, if
$f_s=1/2\pi$ (i.e. a flat uniform distribution), then $\psi=0$ as well. Thus, we would reject
uniformity if $\psi$ exceeds a chosen positive critical value.

It was noted by DSR that by replacing $f(\theta)$ with a density estimator $\hat{f}(\theta)$
based on the observed folded phases $\theta_i$, we retrieve test statistics specific to the kernel of
the density estimator. 
Selecting the Fourier series
estimator $\hat{f}_m$ with $m$ harmonics (see Eqn (5) of DSR) as representative of the light curve, 
results in the well-known $Z^2_m$ given the proper normalisation (Eqn (7) of DSR) 
\begin{equation}
 Z^2_m=2\pi N \psi(\hat{f}_m).
\end{equation}
Since $\psi(f)$ scales with $p^2$ (Eqn \ref{beran}), it is then clear that the statistic $Z^2_m$ should scale as $p^2N$. The quantity $X=p\sqrt{N}$ is then also the approximate Gaussian significance of the point source on the skymap if the background level is well known.

The main problem raised by DSR is that we do not know a-priori the optimal number of harmonics to select. In the case of the popular $Z^2_m$ test, the optimal number of harmonics would depend on $X$ and the pulse profile shape. 

The rationale behind the {\it H-test} was to find a consistent estimator for $f(\theta)$ where the
number of harmonics $m$ is not chosen 
subjectively by the observer, since selecting a number of $m$ values for $Z^2_m$, until a pleasing result is obtained, may
lead to a false detection, given the difficulty to keep track of the number of trials involved.

Hart (\cite{hart}) derived a technique to obtain the optimal number of harmonics $M$ such that the 
mean integrated squared error ({\it MISE}) between the Fourier series estimator $\hat{f}_M(\theta)$ and the
true unknown light curve shape is a minimum (see DSR for a review). Thus, $\hat{f}_M(\theta)$
would give the best Fourier series representation of the true pulse profile, given the constraints
imposed by the statistics and inherent pulsed fraction. Since the minimum of the {\it MISE} involves the finding of
a maximum quantity over all harmonics $m$, DSR redefined this maximal (optimal) quantity in terms of the
so-called {\it H-test} statistic (named after Hart):
\begin{equation}
 H\equiv \max_{1\le m \le 20}(Z^2_m-4m+4)=Z^2_M-4M+4\ge 0.
\end{equation}
Uniformity would correspond to low values of $M$ and also low values of $H$, whereas large
values of $Z^2_M$ (relatively strong pulsed signal) corresponds to large values of $H$. 
Subtracting $4M$ and adding 4 (to ensure 
positive values of $H$ only) effectively limits the variance of $H$ in the absence of a pulsed signal.
It is clear that the $H$ test represents a rescaled version of the $Z^2_m$ test: if $M=1$, then
we retrieve the well-known {\it Rayleigh test}.
DSR have shown that this test is powerful against a wide range of alternatives,
which makes this test attractive if the pulse profile shape is not known {\it a priori}.

To give an example of how $H$ and $M$ depend on the abovementioned significance $X=p\sqrt{N}$
given real data, we selected archival photon arrival times from the public Fermi-LAT data 
base from the Vela pulsar direction, with events selected according to the Fermi-LAT point spread function, but with energies $>500$ MeV. Folding these phases with the given contemporary spin parameters of Vela reveals a pulsed fraction of $p\sim 0.96$, i.e. nearly 100\% and from $N=600$ events
(4 days integration)
we already obtain $M=20$. By adding randomly generated events (i.e. uniformly distributed pulse phases) to the signal events we effectively reduce $p$ and hence $X$, so that $H$ and $M$ should also decrease accordingly. Figure 1 shows how $H$ and $M$ relates to $X$: For $X>3$ (stronger than a 
$\sim 3 \sigma$ signal) we see that $H$ scales with $X^2$, with $H=1.9X^2$, whereas the 
optimal $M$ is already $>10$ for $X>7$. 

\begin{figure}
   \centering
  \includegraphics[width=8 cm]{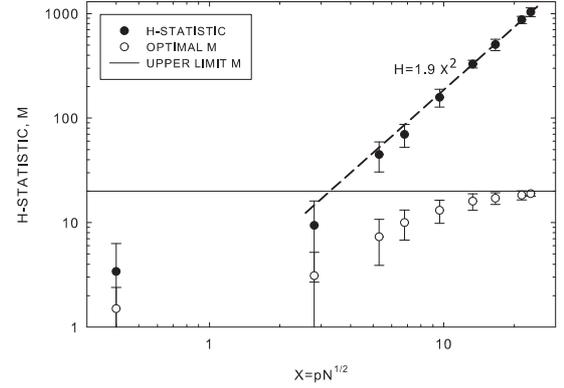}
        \caption{The dependence of $H$ and $M$ on the approximate skymap significance $X$ for
$\gamma$-ray pulse phases above 500 MeV from the Vela pulsar as described in the text.
The errors represent standard deviations based on 20 independent Fermi-LAT Vela pulsar data sets, each of length up to $\sim 4$ days. The horisontal line represent the upper limit of $M=20$, whereas the dashed line represent a fit of the form $H=1.9X^2$ for $X>3$.}
	\label{fig1}
\end{figure}

This figure is also quite useful to see what typical $H$ and $M$ values we may expect for a typical Vela-like pulsar if we know the strength of the signal as derived from a point source on the skymap, and assuming the excess is due to such a pulsed signal.

\section{The revised probability distribution of $H$ for uniformity}

The calculations were performed on the Institutional Cluster of the North-West 
University, Potchefstroom campus. To parallelize the computations, we used the 
RngStream package (L'Ecuyer \cite{ple02}), which guarantees independent, non-overlapping 
substreams of random numbers. The repitition cycle for this random number generator is
$3\times 10^{57}$, which is certainly large enough for our purposes. A total of $4\times10^{14}$ samples were 
calculated in this way.

In the case of large statistics ($N>100$) we do not need to simulate individual arrival times directly
(see the approximate correction factors in Figures 1 and 2 of DSR in the case of low
statistics - $N<100$), so that we only need to simulate the $Z^2_m$ statistics directly:
Since $Z^2_m$ is the sum of $m$ $\chi^2_2$-statistics, we can simulate a $\chi^2_2$ statistic
directly from a uniformly distributed random number $r\in [0,1)$ by taking the transformation $-2\ln {r}$ and adding these
numbers to give $Z^2_m$. This speeds up the process considerably.
A total of $4\times 10^{14}$ values of $H$ were simulated in this way and the results
are shown in Fig. 2. The distribution is everywhere consistent with an exponential
distribution with parameter $\lambda=0.398$ (or 0.4), except for $H>70$ where a downturn
relative to the $0.4$ index is possibly seen. 
DSR arrived at the same
precise value of $\lambda=0.398$ for small values of $H$ (i.e. less than 23)
since the random number generator used by DSR did not yet reach the limit of its random cycle
for the number of simulations required to reach $H\sim 23$ (with a relatively small error on the 
corresponding probability) and should therefore reveal the same 
result as ours for $H<23$.

It is thus clear that the probability distribution of the {\it H}-statistic 
can be conservatively described by the simple formula
\begin{equation}
 Prob(>H)=\exp{(-0.4H)}.
\end{equation}

\section{Incoherent Stacking}

Suppose we analysed the data from $K$ pulsars, or, $K$ independent observations
of the same pulsar, where the effect of e.g. unrecorded glitches excluded the
possibility of analysing all the data as one single coherent set of arrival times.
In this case we want to see if there is a net signal represented by $K$ values of the 
{\it H}-statistic. Suppose we arrive at a set of $K$ such values of $H$, given by
$H_i$, $i=1, ...,K$.

Since we have shown (to the probability level of $\sim 10^{-14}$) that the
$H$-statistic follows that of an exponential distribution with parameter $\lambda=0.4$,
we can stack these test statistics through the sum
\begin{equation}
 H_T=\Sigma_{i=1}^K H_i,
\end{equation}
which is known to follow the Erlang-K distribution with parameter $\lambda$, so that
the significance (or probability for uniformity) of such stacking is given by the simple analytical expression 
(see e.g., Leemis \& McQueston \cite{leemis} on univariate distribution relationships, which includes the Erlang distribution
as the sum of independent exponential variables with parameter $\lambda$.)
\begin{equation}
 P(>H_T|K,\lambda)=\Sigma_{n=0}^{K-1}\frac{\exp{(-\lambda H_T)}(\lambda H_T)^n}{n!}.
\end{equation}

\section{Conclusions}

For $M=1$ we retrieve the well-known {\it Rayleigh test}, with the exception that the parameter for
the exponential distribution has been reduced from $\lambda=0.5$ (for the {\it Rayleigh test})
to $\lambda=0.4$ for the $H-test$. This slight loss of sensitivity is the effect of the 
number of trials taken implicitly into account as a result of the search through $m$ within the $H-test$. 

Finally, it is clear that the corrected distribution of the {\it H}-statistic follows a simple
exponential with parameter $\lambda=0.4$ and evaluation of results for $H>23$ (i.e. the $10^{-4}$ significance
level) would yield more significant results compared to the old distribution presented by DSR.
For example, for $H=50$ a probability of $4\times 10^{-8}$ is typically quoted in the literature,
whereas the true probability for uniformity is actually $2\times 10^{-9}$ - already a factor 20 smaller.

A Fermi-LAT example of the Vela pulsar ($>500$ MeV) shows clearly values for $M$ up to 20 for ``skymap'' significances $X=p\sqrt{N} > 20$, whereas $H$ scales with $H\propto X^2$ as expected for Beran-type tests. The scaling $H=1.9X^2$ can be used to predict $H$-test statistics for Vela-like pulsars above 500 MeV if we assume the excess on the skymap is all pulsed.

The hard tail of the distribution beyond $H>23$ presented by DSR probably arose from the repitition cycle of
the random number generator used in those days, so that the same fluctuations at large $H$ values were repeated
given the finite cycle length of the generator used. In this case we however used a generator with a cycle time
much longer than $10^{14}$, in which case we did not see the repitition of outliers as a result of a 
finite cycle length. Confirmation of the possible break (i.e. downturn) in the probability distribution at $H>70$ requires extensive simulations beyond $4\times 10^{14}$ and is beyond the scope of this paper.
 
\begin{figure}
   \centering
  \includegraphics[width=8 cm]{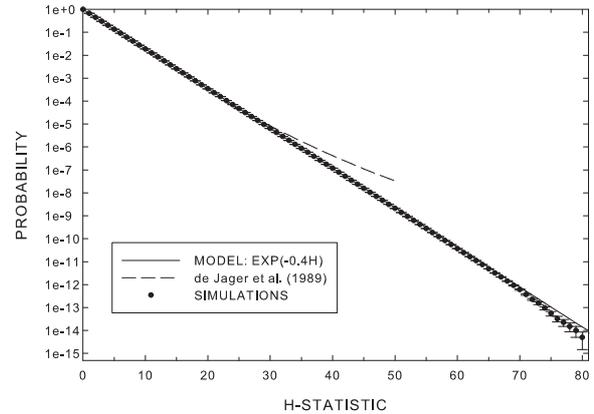}
        \caption{The distribution of the {\it H-statistic} derived from $4\times 10^{14}$ Monte Carlo simulations
with the best fit model for the cumulative probability $P(>H)=\exp{(-0.4H)}$ (solid line) and
the old version of the probability distribution given by DSR shown as a dashed line.}
	\label{fig2}
\end{figure}

\begin{acknowledgements}
   The authors are
  grateful for partial financial support granted to them by the South African National Research Foundation (NRF) and by the Meraka Institute as part of the funding for the South African Centre for High Performance Computing (CHPC).
\end{acknowledgements}


\begin{thebibliography}{}
 
 \bibitem[2009]{fermi1} Abdo, A.A. et al. 2009, Science, 325, (5942), 840

\bibitem[2006]{atwood} Atwood, W.B., Ziegler, M., Johnson, R.P., Baughman, B.M. 2006, ApJ, 652, L49.

 \bibitem[2010]{fermi2} Weltevrede, P., Abdo, A.A., et al. 2010, ApJ, 708, 1426

   \bibitem[1969]{beran} Beran, R.J. 1969, Ann. Math. Stat., 40, 1196

   \bibitem[1989]{bd} Buccheri, R., \& de Jager, O.C. 1989, NATO A.S.I. Workshop on 
Timing Neutron Stars, eds. H. \"Ogelman \& E.P.J. van den Heuvel, Kluwer, Dordrecht, p. 95

   \bibitem[1989]{dsr} de Jager, O.C., Swanepoel, J.W.H., \& Raubenheimer, B.C. 1989, A\&A, 221, 180 (DSR)

  \bibitem[1985]{hart} Hart, J.D. 1985, J. Statist. Comput. Simul., 21, 95

\bibitem[2008]{leemis} Leemis, L.M. \& McQueston, J.T. 2008, American Statistician, 62, 45

  \bibitem[2002]{ple02}L'Ecuyer, P., Simard, R., Chen, E. J., Kelton, W. D. 2002, An 
Object-Oriented Random-Number Package With Many Long Streams and Substreams, Operations Research 50 (6)

\end{thebibliography}

\end{document}